\documentclass{iau}
\usepackage{hyperref}
\usepackage{graphicx}
\usepackage{txfonts}
\usepackage{amssymb}
\usepackage{hyperref}

\title[Mining Sky Surveys with PhotoRApToR] 
{Data-Rich Astronomy: \\Mining Sky Surveys with PhotoRApToR}

\author[Stefano Cavuoti, Massimo Brescia \& Giuseppe Longo]   
{Stefano Cavuoti$^1$, Massimo Brescia$^1$
 \and Giuseppe Longo$^2$}

\affiliation{$^1$INAF - Astronomical Observatory of Capodimonte,\\ I-80131, Napoli, Italy \\ email: {\tt cavuoti@na.astro.it} \\[\affilskip]

$^2$Dept. of Physics, Naples University, \\ Box
I-80126 Napoli, Italy }

\pubyear{2014}
\volume{306}  
\pagerange{119--126}
\jname{Statistical Challenges in Cosmology in the 21st Century}
\editors{A.C. Editor, B.D. Editor \& C.E. Editor, eds.}
\begin{document}

\maketitle

\begin{abstract}
In the last decade a new generation of telescopes and sensors has allowed the production of a very large amount of data and astronomy has become a data-rich science. New automatic methods largely based on machine learning are needed to cope with such data tsunami. We present some results in the fields of photometric redshifts and galaxy classification, obtained using the MLPQNA algorithm available in the DAMEWARE (Data Mining and Web Application Resource) for the SDSS galaxies (DR9 and DR10).
We present PhotoRApToR (Photometric Research Application To Redshift): a Java based desktop application capable to solve regression and classification problems and specialized for photo-z estimation.
\end{abstract}


Everyone who has used neural methods to produce photometric redshift evaluation knows that, in order to optimize the results in terms of features, neural network architecture, evaluation of the internal and external errors, many experiments are needed. When coupled with the needs of modern surveys, which require huge data sets to be processed, it clearly emerges the need for a user friendly, fast and scalable application. This application needs to run client-side, since a great part of astronomical data is stored in private archives that are not fully accessible on line, thus preventing the use of remote applications, such as those provided by the DAMEWARE tool, \cite{brescia2014b} and references therein.
The code of the application was developed in Java language and runs on top of a standard Java Virtual Machine, while the machine learning model was implemented in C++ language to increase the core execution speed. Therefore different installation packages are provided to support the most common platforms.
Moreover, the application includes a wizard, which can easily introduce the user through the various functionalities offered by the tool.

\begin{figure}[ht!]
\centering
\includegraphics[width=9cm]{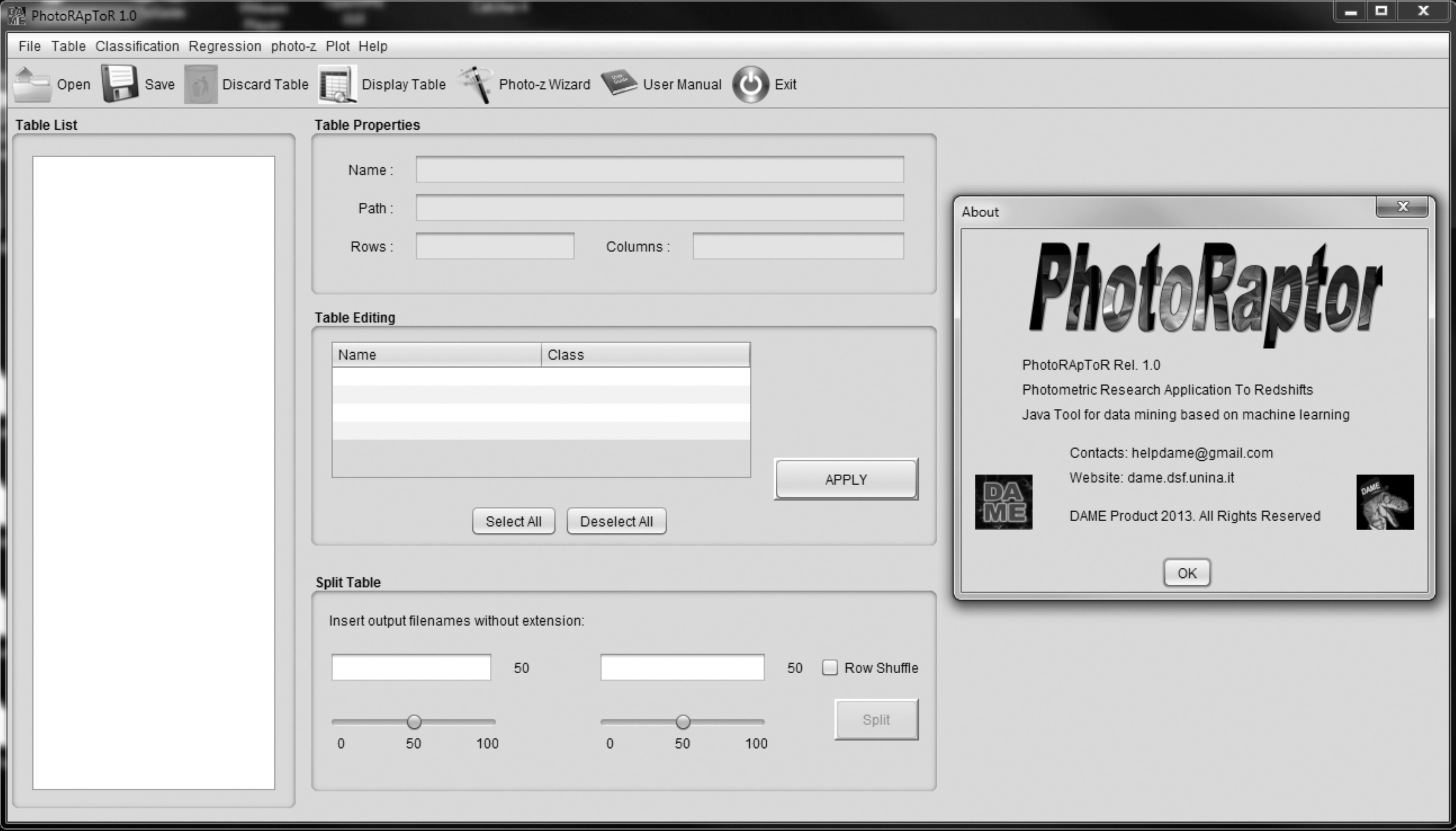}
\caption{The PhotoRApToR main window.}
\label{fig:main}
\end{figure}

The Fig.~\ref{fig:main} shows the main window of the program.

The main features of PhotoRApToR can be summarized as it follows:

\begin{itemize}
\item \textit{Data table manipulation}. It allows the user to navigate throughout his/her data sets and related \textit{metadata}, as well as to prepare data tables to be submitted for experiments. It includes several options to perform the editing, ordering, splitting and shuffling of table rows and columns. A special set of options is dedicated to the missing data retrieval and handling, for instance Not-a-Number (NaN) or not calculated/observed parameters in some data samples;
\item \textit{Classification experiments}.
The user can perform general classification problems, i.e. automatic separation of an ensemble of data by assigning a common label to an arbitrary number of their subsets, each of them grouped on the base of a hidden similarity.
The classification here is intended as \textit{supervised}, in the sense that there must be given a subsample of data for which the right label has been previously assigned, based on the \textit{a priori} knowledge about the treated problem. The application will learn on this known sample to classify all new unknown instances of the problem see for instance \cite{brescia2012b, cavuoti2014};
\item \textit{Regression experiments}. The user can perform general regression problems, i.e. automatic learning to find out an embedded and unknown analytical law governing an ensemble of problem data instances (patterns), by correlating the information carried by each element (features or attributes) of the given patterns. Also the regression is here intended in a \textit{supervised} way, i.e. there must be given a subsample of patterns for which the right output is \textit{a priori} known. After training on such Knowledge Base, the program will be able to apply the hidden law to any new pattern of the same problem in the proper way, see for instance \cite{brescia2014, cavuoti2012, brescia2013};
\item \textit{Photo-z estimation}. Within the \textit{supervised} regression functionality, the application offers a specialized toolset, specific for photometric redshift estimation. After the training phase, the system will be able to predict the right photo-z value for any new sky object belonging to the same type (in terms of input features) of the Knowledge Base;
\item \textit{Data visualization}. The application includes some $2D$ and $3D$ graphics tools, for instance multiple histograms and multiple $2D$/$3D$ scatter plots. For instance, such tools are often required to visually analyze and explore data distributions and trends;
\item \textit{Data statistics}. For both classification and regression experiments, a statistical report is provided about their output. In the first case, the typical confusion matrix \cite{stehman1997} is given, including related statistical indicators such as classification efficiency, completeness, purity and contamination for each of the classes defined by the specific problem. For what the regression is concerned, the application offers a dedicated tool, able to provide several statistical relations between two arbitrary data vectors (usually two columns of a table), such as average (bias), standard deviation ($\sigma$), Root Mean Square (RMS), Median Absolute Deviation (MAD) and the \textit{Normalized} MAD, NMAD \cite{hoaglin1983}, the latter specific for the photo-z quality estimation, together with percentages of \textit{outliers} at the common threshold $0.15$ and at different multiples of $\sigma$ \cite{brescia2014, ilbert2009}.\\
\end{itemize}

\begin{figure}[ht!]
\centering
\includegraphics[width=7cm]{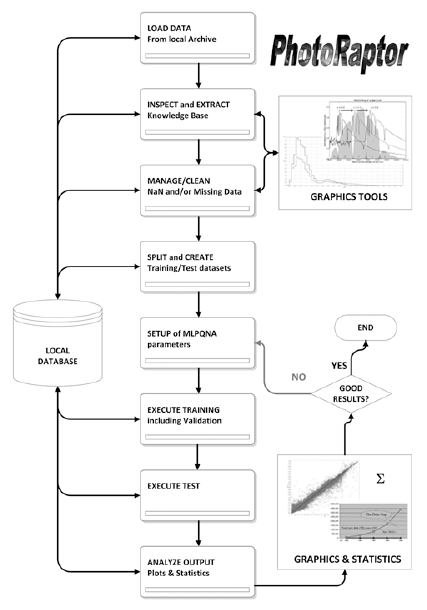}
\caption{The workflow of a generic experiment performed with PhotoRApToR.}
\label{fig:workflow}
\end{figure}

In Fig.~\ref{fig:workflow} the layout of a general PhotoRApToR experiment workflow is shown. It is valid for either regression and classification cases.
The application is available for download from the DAME program web site:\\
\indent \url{http://dame.dsf.unina.it/dame_photoz.html\#photoraptor}\\

\textbf{Perspective and conclusions}\\
Nowadays the technological evolution of  astronomical detectors and instruments has been so fast to render physically impossible to move the data from their original repositories. This implies that, as it has always been asked for but never implemented, we must be able to move the programs and not the data. Therefore, the future of any data-driven service depends on the capability and possibility of moving the data mining applications to the data centers hosting the data themselves.
In this scenario PhotoRApToR rapresents our test bench of a desktop application prototype, fine tuned to tackle specific astrophysical problems in the big data era.

\end{document}